\documentclass[12pt,a4paper]{article}

\usepackage{graphicx,color,tikz}
\usepackage{mathtools}

\begin{document}

\title{The Mathematics of Market Timing} \author{Guy
  Metcalfe\thanks{guy.metcalfe@monash.edu, ORCID ID
    0000--0003--4679--8663} \\ School of Mathematical Sciences \\
  Monash University \\ Australia} \date{12 December 2017}

\maketitle


\abstract{Market timing is an investment technique that tries to
  continuously switch investment into assets forecast to have better
  returns.  What is the likelihood of having a successful market
  timing strategy?  With an emphasis on modeling simplicity, I
  calculate the feasible set of market timing portfolios using index
  mutual fund data for perfectly timed (by hindsight) all or nothing
  quarterly switching between two asset classes, US stocks and bonds
  over the time period 1993--2017.  The historical optimal timing path
  of switches is shown to be indistinguishable from a random sequence.
  The key result is that the probability distribution function of
  market timing returns is asymetric, that the highest probability
  outcome for market timing is a below median return.  Put another
  way, simple math says market timing is more likely to lose than to
  win---even before accounting for costs.  The median of the market
  timing return probability distribution can be directly calculated as
  a weighted average of the returns of the model assets with the
  weights given by the fraction of time each asset has a higher return
  than the other.  For the time period of the data the median return
  was close to, but not identical with, the return of a static 60:40
  stock:bond portfolio.  These results are illustrated through Monte
  Carlo sampling of timing paths within the feasible set and by the
  observed return paths of several market timing mutual funds.}


\newpage

\section{Introduction}

Market timing is an investment technique whereby an investment manager
(professional or individual) attempts to anticipate the price movement
of asset classes of securities, such as stocks and bonds, and to
switch investment money away from assets with lower anticipated
returns into assets with higher anticipated returns.  Market timing
managers use economic or other data to calculate propitious times to
switch.  Market timing seems a popular approach to investment
management, with Morningstar listing several hundred funds in its
tactical asset allocation (TAA) category---TAA being an industry name
for market timing---and mainstream fund managers advertising their
ability to switch to defensive assets when stock markets seem poised
for a downturn.  The antithesis of market timing, and another broadly
popular investing approach, is buy-and-hold, whereby investment
managers allocate static fractions of their monies to the available
asset classes and then ignore market price gyrations.


Is market timing likely to be successful relative to investing in a
static allocation to the available asset classes?  The literature in
this area is focused on developing sophisticated statistical tools
that can detect and measure the market timing ability of professional
fund managers \cite{Henriksson_timing_1984}.  Numerous uses of these
techniques over decades have produced mixed results
\cite{Bello_timing_1997,Becker_timing_1999,Goetzmann_daily_2000,Chance_timers_2001,Jiang_time_2007,Ptak_practice_2012}.
Some authors detect no market timing ability, while others report
statistically significant evidence of market timing ability.  On the
other hand, Dalbar measures the market timing results of the average
individual investor through mutual fund sales, redemptions and
exchanges \cite{Dalbar_QAIB_2016}.  These studies find unambiguously
that market timing by the average investor is unsuccessful relative to
a static allocation.  The ambiguous results for successful market
timing from professional managers suggests that, at minimum, it is
difficult to market time successfully, while the unambiguous results
for individuals strongly suggests that it is easy to market time
unsuccessfully.



My goal here is both different and simpler than statistical tests to
detect market timing.  I want to create a simple model to ask the
question, what is the likelihood of successful market timing?  Or more
precisely, what is the return probability distribution function (PDF)
for market timing?  Is the PDF of market timing returns symmetric?  If
it is hard to obtain above average returns by market timing, is it
also hard to obtain below average returns?  What is the most basic
mathematics of market timing?


I try in this paper to evoke a similar spirit to Sharpe's "The
Arithmetic of Active Management" \cite{Sharpe_arithmetic_1991}, in
which elementary arithmetic is all that is required to demonstrate why
active management must in aggregate under perform low-cost index
funds.  While I will need to invoke elementary probability theory, it
will show that the most probable outcome of market timing is to under
perform a buy-and-hold, suitably weighted average of the available
asset classes.  Moreover, as I build the simple model from the returns
of US stock and bond total market index funds since 1993, market
returns over that time period mean that the suitably weighted average
portfolio, while not identical with the 60:40 stock:bond balanced
fund, is in practice barely distinguishable from it.



In the rest of the paper my approach will be to calculate the
boundaries of the feasible set of market timing portfolios using fund
data for perfectly timed (by hindsight) switching between two asset
classes, stocks and bonds.  From this analysis I also obtain the
historically optimal timing path of switches, which the
NIST\footnote{National Institute of Standards and Technology, U.S.\/
  Department of Commerce, \texttt{www.nist.gov}.} suite of tests for
randomness shows is indistinguishable from a random sequence.  The key
elementary result is that the geometric mean of market timing returns
has an asymmetric PDF.  One implication of this is that the most
probable market timing return is below the median return, which can be
directly calculated to be given by a static portfolio weighted by the
relative fraction of time periods that each asset class outperforms
the other.  These results are illustrated through Monte Carlo sampling
of timing paths within the feasible set and by the return paths of
several market timing funds with comparably long, publicly available
data.  To begin, in the next section I describe the data.


\section{Data}
\label{sec:data}

\begin{figure}
  \centering
  \includegraphics[width=0.75\textwidth]{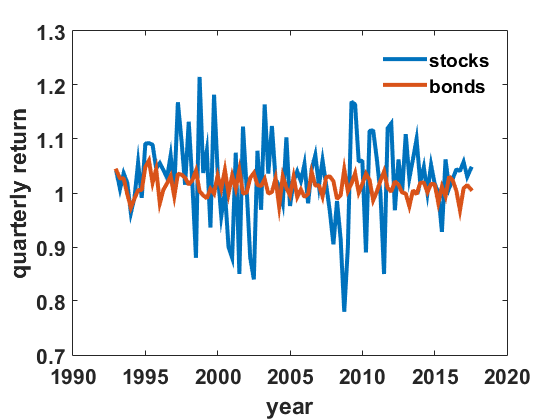}
  \caption{Quarterly return time series for stock and bond total
    market index funds, 1993--2017.  Returns are in multiplicative
    form.}
  \label{fig:quarterlyreturns}
\end{figure}

The data consists of time series of quarterly returns for three index
funds starting in 1993, the advent of the youngest of the three funds,
and ending in Q3 2017.  The series covers 24 years, and there are
$N = 99$ data points per series.  The funds, all from Vanguard, are
Total Stock Market, Total Bond Market, and Balanced Index, the last a
static portfolio of 60\% Total Stock and 40\% Total Bond.  Other
information on these funds is in appendix~\ref{sec:index_funds}.
Figure~\ref{fig:quarterlyreturns} shows the quarterly return time
series for stocks and bonds.  Because the data are from live funds,
calculated return paths are net of management and trading costs;
however, tax consequences are ignored.  For quarterly switching taxes
would likely be substantial, but the effect would only dampen the
spread of net returns and change only the quantitative, not the
qualitative results of the model.  Note that because fund data are the
basic building blocks of the model, all return paths calculated could
have been obtained by an investor during the time period.

Since the way to calculate total return is to multiply the sub-period
returns together, I trivially transform the original data to
multiplicative form, e.g.\/ a $+3$\% return becomes $1.03$ and a
$-3$\% return becomes $0.97$.  The differences between multiplicative
and additive random processes will be important in the subsequent
analysis.


\section{Two Asset, All or Nothing Market Timing Model}
\label{sec:model}

Here I define the simple two asset market timing model with all or
nothing quarterly switches.  Using perfect hindsight, it is easy to
identify the best and worst possible market timing portfolios, which
form the boundaries of the feasible return paths for all market timing
portfolios, i.e.\/ all possible market timing portfolios lie between
the boundaries of the feasible set\footnote{Technically it is all
  market timing portfolios that conform to the assumptions of the
  model; however, in section~\ref{sec:discussion} we will see that
  real, non-conforming market timing funds fall within the feasible
  set.}.  I reveal the optimal (highest possible return) timing
sequence and test it for randomness.  Section~\ref{sec:pdf} focuses on
deriving the return PDF for the model.

\subsection{Model}

The model consists of quarterly all or nothing switches between stocks
and bonds.  In the $i$th time period $t_i$ the return of stocks is
denoted $r_{si}$ and the return of bonds is denoted $r_{bi}$.  A {\em
  timing path} is the binary sequence $f_i$ that is
\begin{equation}
\label{eq:timing_path}  
  f_i =
  \begin{cases}
    1 & \quad \mathrm{if \: during} \quad t_i \quad r_{si} > r_{bi} \\
    0 & \quad \mathrm{if \: during} \quad t_i \quad r_{si} < r_{bi}.
  \end{cases}
\end{equation}
In other words $f$ is set to $f = 1$ when the stock return is larger
than the bond return and set to $f = 0$ when the bond return is larger
than the stock return.  A special class of timing path has
$f = \textrm{constant}$ and is termed a static allocation or
buy-and-hold portfolio.  I call a {\em return path}, denoted $\rho$,
the sequence of returns generated by a particular timing path $f_i$.
The $j$th return path is given by
\begin{equation}
\label{eq:return_path}
\rho_j = \prod_i^N \left( f_{ij} r_{si} + (1 - f_{ij}) r_{bi} \right).
\end{equation}
The geometric mean of a return path is given by $\rho_j^{1/N}$.  

\subsection{Feasible Set}

\begin{figure}
  \centering
\begin{tabular}{cc}  
  \includegraphics[width=0.5\textwidth]{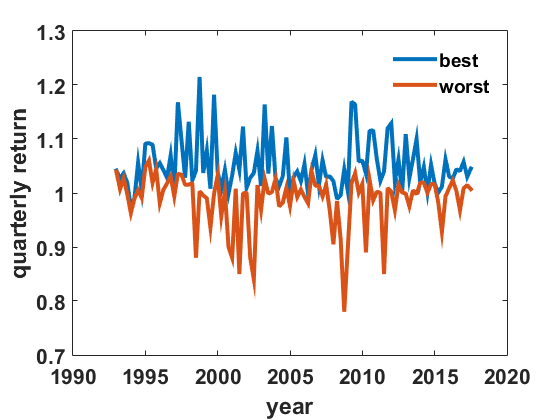} & 
  \includegraphics[width=0.5\textwidth]{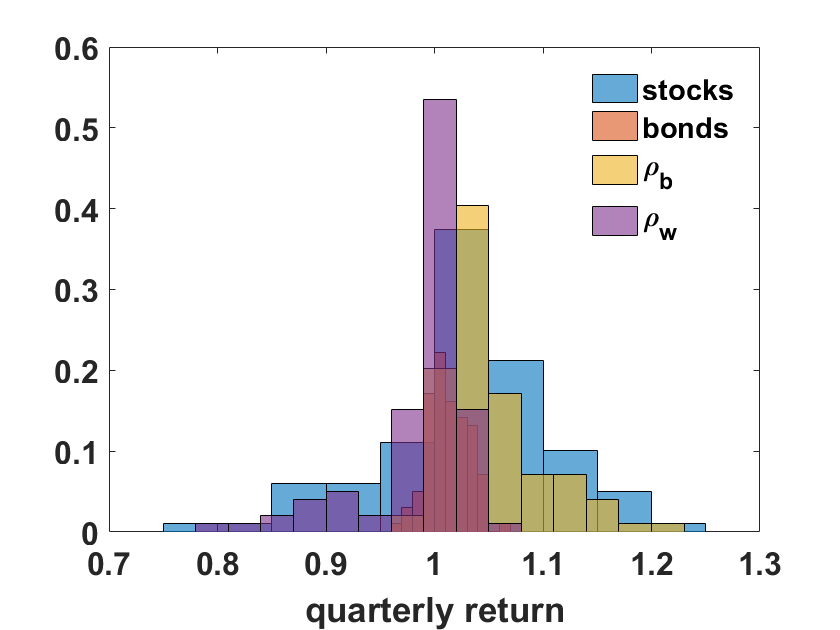} \\
  (a) & (b) \\
  \multicolumn{2}{c}{\includegraphics[width=0.75\textwidth]{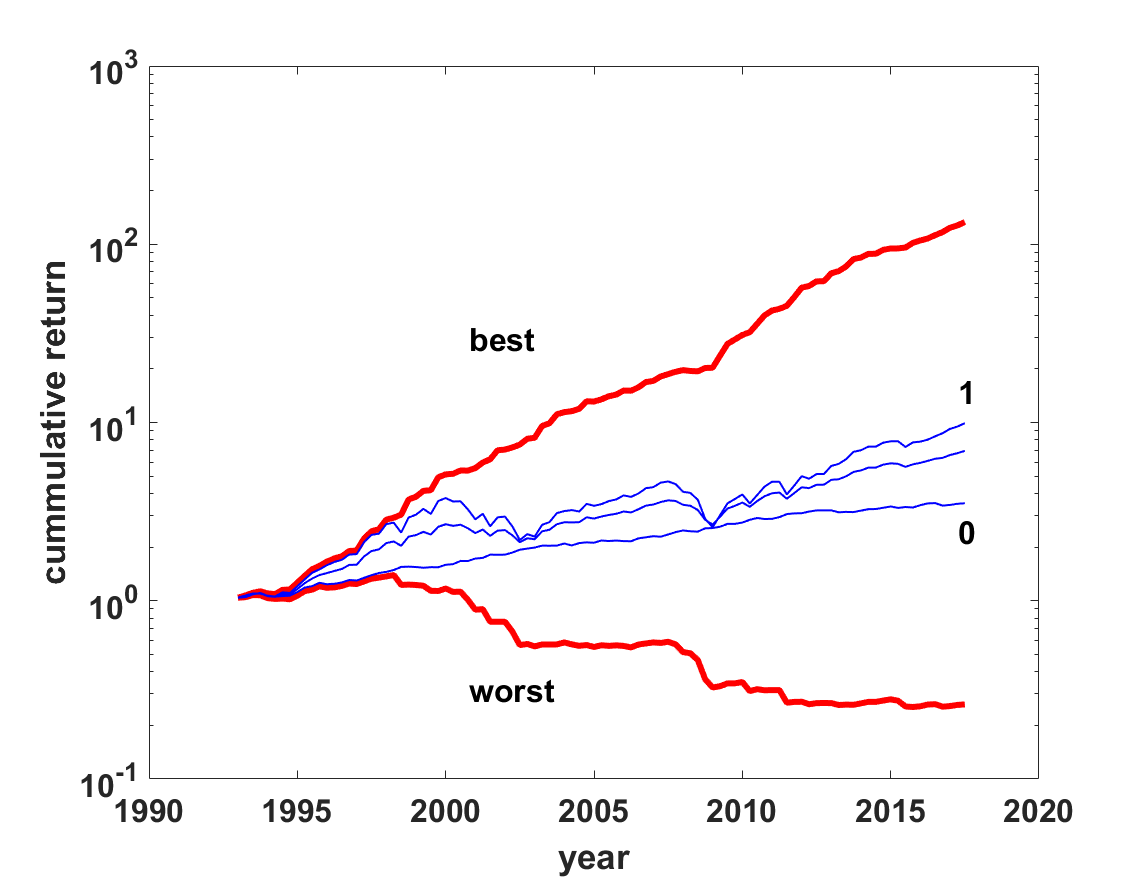}} \\
  \multicolumn{2}{c}{(c)}
\end{tabular}
\caption{Two asset, all or nothing market timing model switches to
  whichever of the two assets classes will have the better return that
  quarter.  (a) Quarterly returns of the best and worst market timing
  portfolios as a function of time in multiplicative form.  (b)
  Histograms of returns for the indicated data sets.  (c) Feasibility
  envelope plotted on semi-log axes.  Thick red lines are the best and
  worst possible return paths over this time period.  Blue lines are
  the three data sets: stocks $(f = 1)$, bonds $(f = 0)$, and balanced
  $(f = 0.6)$.  The fixed portfolio lines order as expected from
  $f = 0$ to $f = 1$.}
  \label{fig:envelope}
\end{figure}

With perfect hindsight the best and worst performing return paths are
easily found.  In the notation of Matlab code\footnote{Matlab code and
  data are available at \\
  \texttt{https://www.dropbox.com/s/6i82p9phq7q56be/timing.m?dl=0}.}
equation~\ref{eq:return_path} becomes for the best $\rho_b$ and worst
$\rho_w$ possible return paths
\begin{subequations}
\label{eq:optimum_paths}
\begin{align}
  \rho_b &= \textrm{cumprod(max(stocks, bonds))} \\
  \rho_w &= \textrm{cumprod(min(stocks, bonds))},
\end{align}
\end{subequations}
and the timing path for $\rho_b$ is given by
$f_b = (\textrm{stocks} > \textrm{bonds})$; similarly for $\rho_w$.
Figure~\ref{fig:envelope}(a) shows the quarterly return series for
$\rho_b$ and $\rho_w$, while figure~\ref{fig:envelope}(b) shows
histograms of quarterly returns for stocks, bonds, $\rho_b$, and
$\rho_w$.  There are no surprises: partitioning returns by
equation~\ref{eq:optimum_paths} puts the positive return, right tail
of the stocks distribution into $\rho_b$, while excluding the negative
return, left tail.  The reverse happens to $\rho_w$.

Figure~\ref{fig:envelope}(c) plots several return paths on semi-log
axes.  The best and worst possible return paths for this period are
thick red lines.  Blue lines are the fund data for stocks $(f = 1)$,
bonds $(f = 0)$ and balanced $(f = 0.6)$.  The returns of the fixed
portfolios are ordered as expected with $f = 0$ producing the lowest
returns of the fixed $f$ portfolios and $f = 1$ producing the highest.
Note, however, that the large difference in returns normally
associated with stocks and bonds is dwarfed by the difference in
returns between the best and worst market timing portfolios.  The
potential reward to successful market timing is clearly enormous;
however, just an enormous is the potential penalty to unsuccessful
market timing.

The best and worst possible return paths demark the feasible set of
return paths for the two asset model.  All possible return paths (all
possible market timing paths $f_i$) fall inside the envelope made by
$\rho_b$ and $\rho_w$.  As the model has all or nothing switches, the
number of possible paths of length $N$ is $2^N$.  As the data set has
$N = 99$, the number of possible return paths is
$2^{99} \sim 10^{29}$, which is large.

\subsection{The Unpredictable Optimal Timing Path}

\begin{figure}
  \vspace*{-4em}
  \centering
  \includegraphics[width=\textwidth]{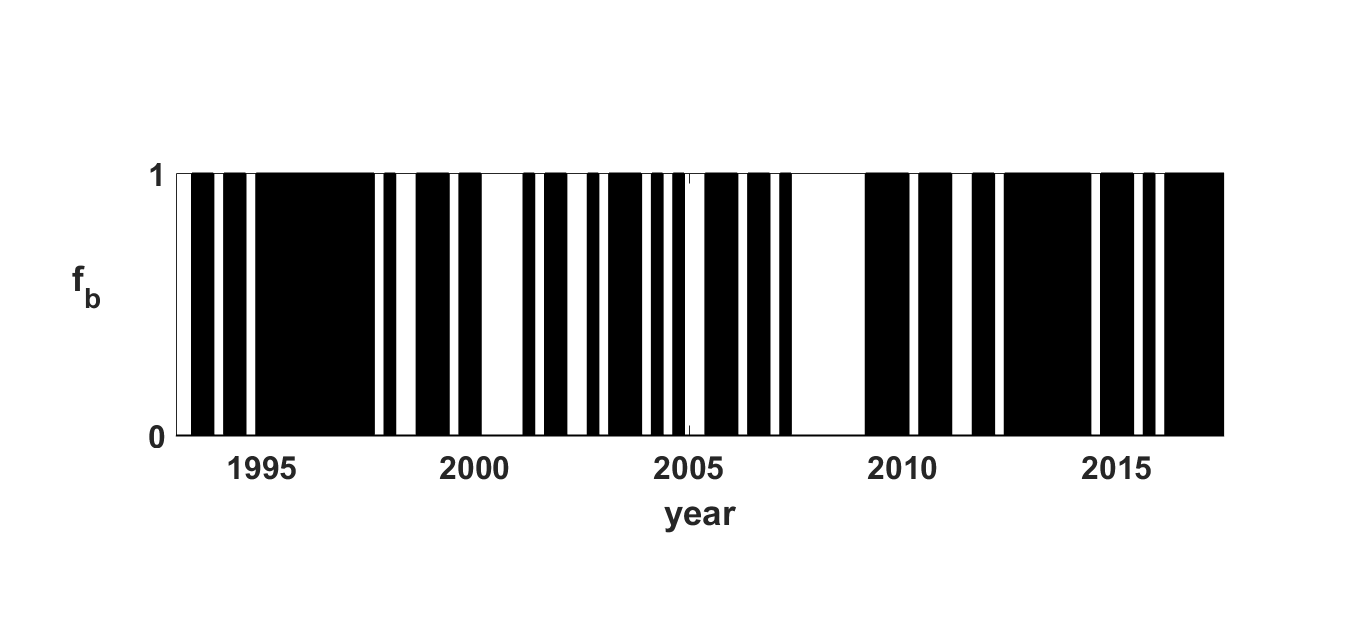}
  \vspace{-4em}
  \caption{Optimal timing path $f_b$ that would have produced the
    highest possible return path $\rho_b$ over the time period.  Black
    regions have $f_i = 1$ (stocks $>$ bonds).  White regions have
    $f_i = 0$ (bonds $>$ stocks).}
  \label{fig:optimal_timing_path}
\end{figure}

Figure~\ref{fig:optimal_timing_path} shows the historical optimal
timing path $f_b$ that produces the highest possible return path
$\rho_b$ over the time period.  Black regions have $f_i = 1$ (stock
return $>$ bond return).  White regions have $f_i = 0$ (bond return
$>$ stock return).  It will be convenient to define $p$ as the
fraction of time periods in which $f = 1$, which is easily calculated
by summing $f_b$ and dividing by $N$.  For this data
$p = p_b \approx 0.64$: over this time period approximately 2/3 of the
time stocks returned more than bonds.  While the optimal timing path
$f_b$ is not random like a coin flip ($p_b \ne 1/2$),
figure~\ref{fig:optimal_timing_path} shows no pattern readily
discernible to the eye.  Is $f_b$ random?



It is worth distinguishing random and unpredictable.  The historically
optimal timing path is not a random bit sequence because ones occur
about two-thirds of the time.  Nonetheless, the important question is
can I predict the next element in the sequence, given knowledge of the
previous elements of the sequence?  How can a sequence be not random
but at the same time unpredictable?  Consider a 6-sided die, of which
four sides have a one and two sides have a zero.  For each fair roll
of the die there is a two-thirds probability of a one and a one-third
probability of a zero.  Since each fair roll of the die is independent
of all rolls that have come before, there is no way to predict from
the past sequence of rolls what the next roll of the die will produce.
The analogy is not perfect because $p_b$ is not known {\it a priori},
and in fact $p_b$ could be different over different time periods.

Leaving the details to appendix~\ref{sec:nist_details}, I use the suit
of 15 tests published by NIST \cite{Bassham_NIST_2010} and designed
for the purpose of verifying random number generators for
cryptography.  While most of the NIST tests, in order to ensure an
accurate test, require orders of magnitude longer bit sequences than
the financial time series provides, for four of the tests the $N = 99$
bit length of $f_b$ is close to the suggested minimum length.  Again,
leaving details to appendix~\ref{sec:nist_details}, the result of
those four tests is that $f_b$ is random (unpredictable) at the 99\%
confidence level.

While the historically optimal timing sequence $f_b$ is clearly
special in some sense---the probability of that particular sequence to
occur is $2^{-99}$---the question is what, if anything, distinguishes
$f_b$ from any other random timing paths?  If we look at $f_b$ and
randomly generated timing paths {\em without knowing which is which},
can we distinguish $f_b$ from the masses of possible timing paths?  If
$f_b$ is random, as the NIST tests say it is, there is nothing to tell
why it is special, which says that it is not special, that just by a
$2^{-99}$ random chance, it was special for this time period and that,
in itself, $f_b$ is unpredictable, i.e.\/ it contains {\em no}
information about any future optimal timing path.

\section{Probability Distribution of Return Paths}
\label{sec:pdf}

As the optimal timing path is indistinguishable from a random
sequence, I review elementary properties of random multiplicative
processes, from which it follows that the highest probability outcome of
market timing is a return less than the median of the PDF of market timing
returns.  The return PDF is estimated by Monte Carlo sampling of
random timing paths.  The median of the return PDF can be directly
calculated as the weighted average of the returns of the assets with
the weights given by the fraction of time each asset has a higher
return than the other.  For the time period covered by the data the
median return was close to the $f = 0.6$ balanced index fund.

\subsection{Monte Carlo}

\begin{figure}[tb]
  \centering
  \includegraphics[width=\textwidth]{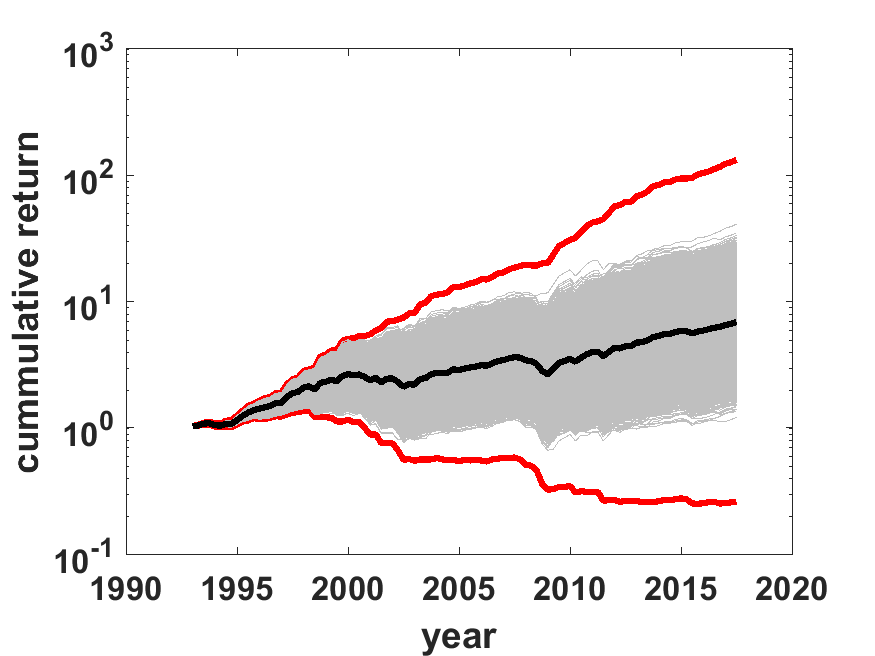}
  \caption{Return paths (gray) for $M = 10^5$ randomly generated
    timing paths.  Red lines are the best and worst market timing
    return paths.  The black line is the observed $f = 0.6$ balanced
    fund returns.}
  \label{fig:paths}
\end{figure}

The distribution of typical returns of the model can be estimated by
Monte Carlo methods. Generate $M$ random timing paths of length $N$
and calculate $M$ return paths with equation~\ref{eq:return_path}.  In
order to match the period data, set random timing paths to have the
same fraction of ones and zeros as the data, i.e.\/ the average value
of $p$ for the $M$ timing paths is set to $p = p_{b}$.  This is done
by using Matlab's \verb|rand| function to generate a length $N$
sequence of random real numbers $n$ drawn from a uniform distribution
in the range $[0, 1]$ and setting each term in the sequence equal to
one if $n < p_{b}$ or to zero if $n \ge p_{b}$.
Figure~\ref{fig:paths} shows $M = 10^5$ return paths as thin gray
lines in a semi-log plot similar to figure~\ref{fig:envelope}(c).  Red
lines are the boundaries of the feasible set, $\rho_b$ and $\rho_w$,
while the thick black line is the data for the $f = 0.6$ balanced
fund.  Before further examination of the return PDF it will be useful
to review several facts about distributions from random multiplicative
processes, such as that of equation~\ref{eq:return_path}.

\subsection{Random Multiplicative Processes}

A {\em sum} of random numbers is guaranteed by the central limit
theorem to converge to a Gaussian (normal) PDF in the limit of a large
number of terms in the sum.  A {\em product} of random numbers, such
as that used in equation~\ref{eq:return_path} to calculate return,
does not share this nice property.  On the contrary, the PDF for a
random multiplicative process (of positive numbers) depends on rare
sequences that generate an asymmetric PDF with a long tail.  The
average value of the PDF (or of any moment) depends sensitively on the
sampling size $M$ and, until $M$ approaches the number of possible
outcomes, becomes larger and larger compared to the mode
\cite{Redner_multiplicative_1990}.

Nonetheless, what can be done is to take the log of the geometric mean
of equation~\ref{eq:return_path} to change the product of returns to a
sum of the log returns:
\begin{equation}
\label{eq:log_geo_mean}
\log\left(\rho_j^{1/N}\right) =
   N^{-1} \sum_i^N \log\left( f_i r_{si} + (1 - f_i) r_{bi} \right).
\end{equation}
Equation~\ref{eq:log_geo_mean} says that the log of the geometric mean
is given by the average of the log return.  The PDF of log return then
does obey the central limit theorem to converge to a Gaussian PDF.
Moreover, if the log of something is distributed as a Gaussian, then
the something has a log-normal PDF \cite{Redner_multiplicative_1990}.
In other words, the return PDF for market timing is log-normal, as a
simple consequence of elementary properties of the logarithm.
Further, if $\mu$ and $\sigma$ are respectively the median and
variance of the Gaussian PDF, then $e^\mu$ is the median and
$e^{\mu - \sigma^2}$ is the mode of the log-normal PDF: the mode,
which is the most probable outcome, is less than the median of the
log-normal PDF.  Thus from elementary considerations the most probable
outcome from market timing is a return that is less than the median of
the return PDF.

\begin{figure}[p]
  \centering
\begin{tabular}{c}
  \includegraphics[width=\textwidth]{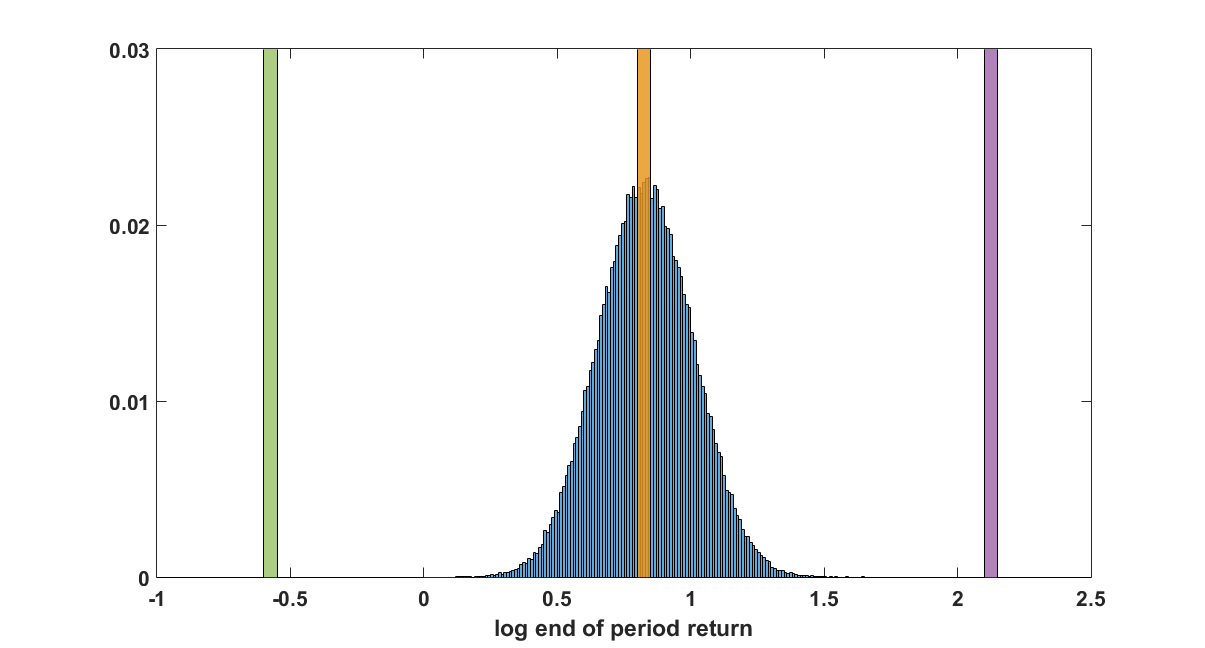} \\
  (a) \\
\begin{tikzpicture}[      
        every node/.style={anchor=south west,inner sep=0pt},
        x=1mm, y=1mm,
      ]   
     \node (fig1) at (0,0)
       {\includegraphics[width=\textwidth]{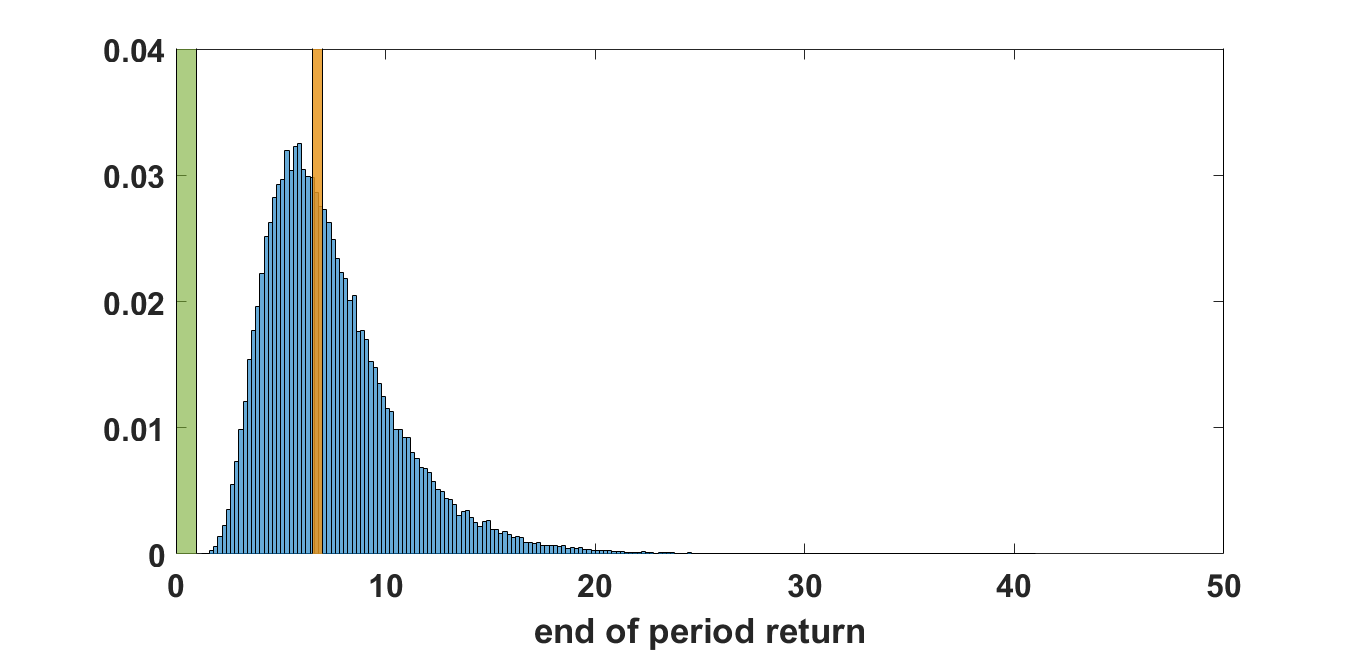}};
     \node (fig2) at (41,21)
       {\includegraphics[width=0.6\textwidth]{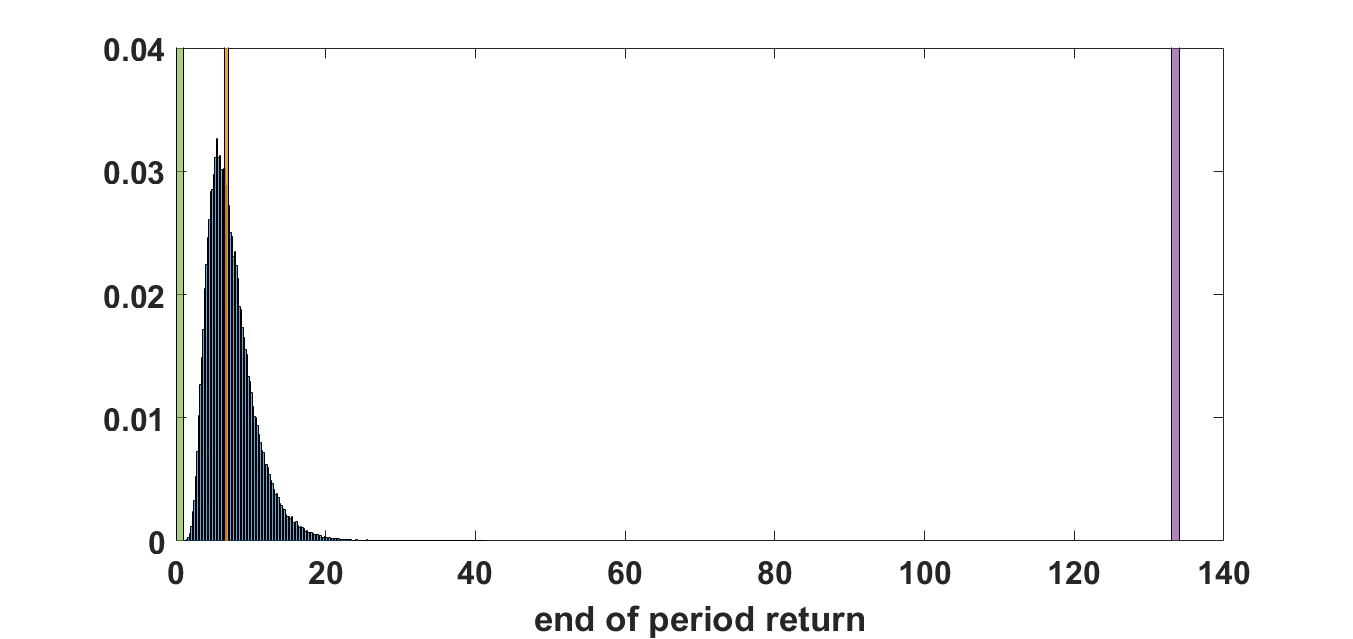}};  
  \end{tikzpicture} \\
  (b)
\end{tabular}
\caption{Probability distribution function of (a) log-return and (b)
  return estimated from $M = 10^5$ trials with
  $p = p_{b} \approx 0.64$.  Green and purple vertical bars are
  respectively the worst and best timing portfolios.  The orange bar
  is the median of the PDF and the observed return of the $f = 0.6$
  balanced index fund, which so closely approximates the median as to
  be indistinguishable at this scale.  Inset of (b) is the full data
  range, showing the extreme low probability position of the optimum
  timing portfolio (purple bar) in the tail of the distribution.}
\label{fig:returnpdf}
\end{figure}

To illustrate, figure~\ref{fig:returnpdf}(a) plots the histogram of
end of period log returns from the Monte Carlo data of
figure~\ref{fig:paths}.  Even though $M = 10^5$ grossly under samples
the order $10^{29}$ distinct paths in the feasible set, convergence to
a Gaussian PDF is evident, as predicted by the form of
equation~\ref{eq:log_geo_mean}.  The green and purple bars at the
extremes are the results for respectively $\rho_w$ and $\rho_b$.  The
orange bar marks the median log-return and the log -return for the
$f = 0.6$ balanced index fund, which are indistinguishable in this
plot, and the reason for this will be discussed in the next section.
Figure~\ref{fig:returnpdf}(b) plots the histogram of the end of period
return (not log-return).  The predicted log-normal form with a long
tail is also evident.  The inset shows the entire data range to
indicate how long the return tail is.  Colored bars have the similar
meaning as in figure~\ref{fig:returnpdf}(a), just for the return PDF
instead of the log return PDF.  The highest probability outcome is the
mode (maximum) of the distribution, which is less than the median
return marked by the orange bar.

\subsection{Expectation Value of the Median}

The expectation value operator $\boldmath E$ gives the most probable
value of a PDF.  After a calculation given in detail in
appendix~\ref{sec:calculation_detailed}, the expectation value of
equation~\ref{eq:log_geo_mean} for the median $\mu$ of the log return
distribution is
\begin{equation}
\label{eq:evalue_log_geo_mean}
\mu = {\boldmath E}\left[\log\left(\rho_j^{1/N}\right)\right] =
   \log\left( p_b  \bar{r}_{s} + (1 - p_b) \bar{r}_{b} \right), 
\end{equation}
where $\bar{r}_{s,b}$ are the geometric mean returns of the stock and
bond assets.  Recall $p_b$ is the observed fraction of time periods
that the stock return exceeds the bond return.  The median of the
distribution of log returns is given by the log of the weighted
average of the two assets with the weights given by the fraction of
time periods $p$ that each asset's return exceeded that of the other.
The median of the return PDF is $e^\mu$.

Note that because over the time period of the data $p_b \approx 0.64$,
that using the log return for the $f = 0.6$ balanced fund for the
right hand side of equation~\ref{eq:evalue_log_geo_mean} well
approximates the exact result for $\rho_b$, which, of course, cannot
be known {\it a priori\/}.  As noted above, in
figure~\ref{fig:returnpdf} the median return and the return for the
$f = 0.6$ balanced index fund are indistinguishable at the scale of
the plot.

It is important to note that figure~\ref{fig:returnpdf} shows the PDF
for {\em costless} market timing.  In practice, market timing costs
higher than the index fund costs would shift the PDF to the left, but
the boundaries of the feasible set and the median of the PDF would not
shift because they are calculated from fund data, which already
includes the small index funds costs.  In practice what the Monte
Carlo simulation estimates is the lower bound of the most likely
shortfall of market timing to the median return given by the
appropriately weighted static portfolio.

\section{Discussion}
\label{sec:discussion}


\begin{figure}[tb]
  \centering
  \includegraphics[width=\textwidth]{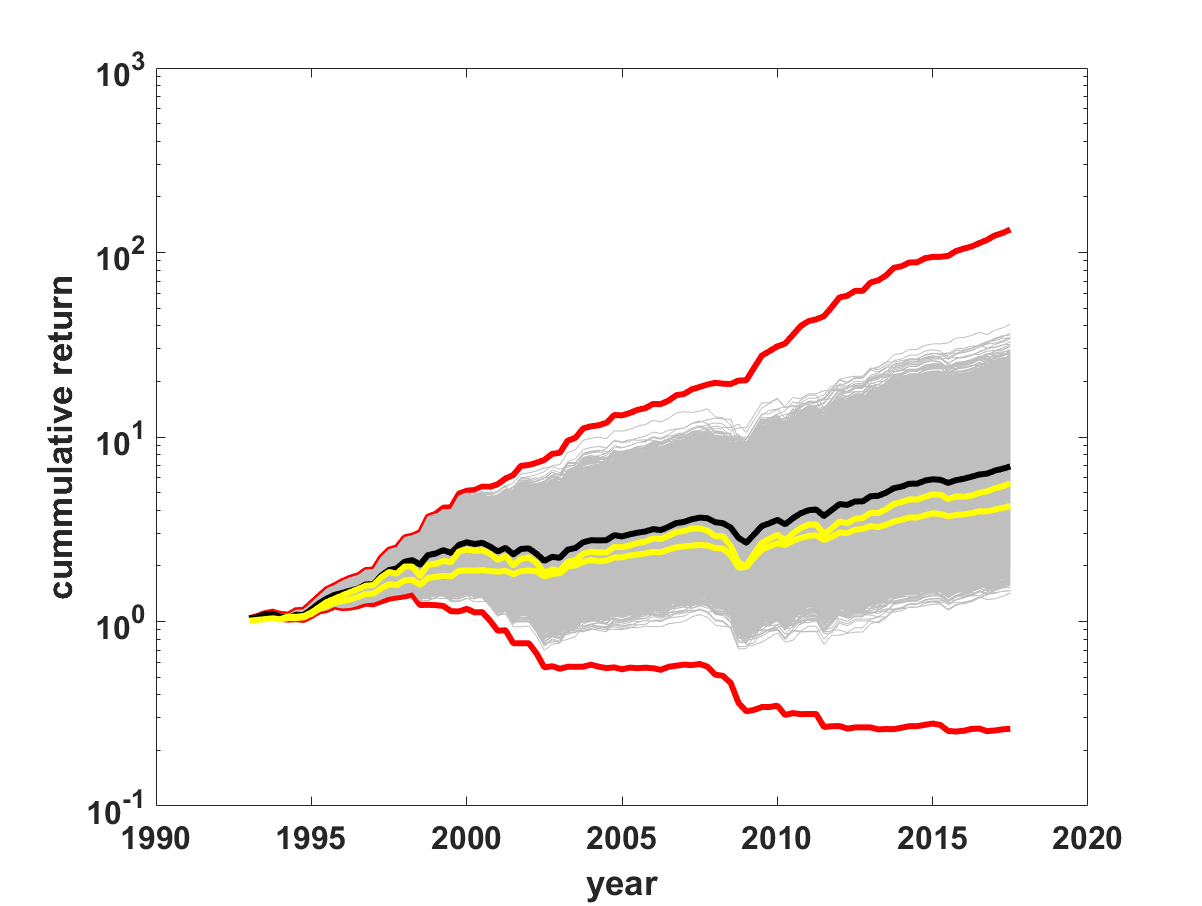}
  \caption{Reprise of figure~\protect\ref{fig:paths} with the addition
    of two market timing funds with publicly available data of
    comparable length (yellow lines).  Red lines are the best and
    worst timing portfolio return paths.  The black line is the
    observed $f = 0.6$ balanced index fund.}
  \label{fig:paths_taa_overlay}
\end{figure}

Several critiques could be leveled at the analysis in this paper.  For
example, adherents of market timing would claim that their timing
systems are not random, therefore they would be able to choose timing
paths to have returns far out on the right tail of the PDF, i.e.\/
that the strategy to generate random paths (random $f$ sequences) is
not representative of actual market timing.  There are two answers to
this.  One is that the feasible set is well-defined and that it is
simply a fact that all market timing paths, {\em no matter how they
  are generated}, are contained in the feasible set.  As such, any
sampling of the feasible set generates valid timing paths.  The second
answer is in figure~\ref{fig:paths_taa_overlay}, which reproduces
figure~\ref{fig:paths} with the addition of the return paths (yellow
lines) of two funds that Morningstar classifies as TAA funds and for
which there are publicly available returns data from 1994, almost as
long as for the index funds data series.  Appendix~\ref{sec:taa_funds}
has details about these two funds, which are rated by Morningstar as
above average.  While these market timing funds were neither limited
to two asset classes, nor did they make all or nothing switches, yet
their return paths are, as expected, contained inside the feasible
set.  The conclusion is that real-life market timers are correctly
characterized---except for costs---by the PDF within the feasible set,
and that random sampling of the PDF does properly characterize the
return distribution expected from market timing schemes.


Figure~\ref{fig:paths_taa_overlay} also nicely illustrates the main
result with live, not simulated, market timing data.  These
long-lived, above average market timing funds trailed the median
return over the time period---and its close proxy, the $f = 0.6$
balanced fund---as simple math says is the most probable outcome.
This longer-term observation is consistent with more recent analysis
covering a much shorter time period but many more TAA funds
\cite{Ptak_practice_2012}\footnote{From \cite{Ptak_practice_2012}, "We
  found that very few tactical funds generated better risk-adjusted
  returns than Vanguard Balanced Index over the extended time period
  we studied.  Not only has the group of tactical allocation funds
  underperformed, but not a single one of them outperformed the
  simple, low-cost, passive fund."}.

A more subtle criticism is that I have not disproved market timing.
This is because of the possibility of hidden variables.  Hidden
variables represent information, such as earnings, book value,
anything, that a market timer could put into a function that produces
a timing path.  While the observed optimal timing path $f_b$ is random
to the extent that it passes the NIST tests, it is possible that there
was a set of hidden variables that could have been combined in a
function that would have produced the optimal timing path $f_b$.  Good
pseudo-random number generators also pass the NIST tests but are
produced by deterministic systems.  Taking into account the fund data
of figure~\ref{fig:paths_taa_overlay}, I think it highly unlikely, but
it could be true and so market timing is not mathematically disproved.
Take comfort in that, dear reader, as you will.

\section{Conclusions}

I have examined a two asset, all or nothing market timing model with
24 years of data from US stock and bond total market index funds from
1993--2017.  The model is deliberately kept simple in order to see the
basic mathematics of market timing at work answering the question,
what is the likelihood of successful market timing?  The boundaries of
the feasible set of market timing paths, within which all market
timing return paths must lie, is easy in hindsight to calculate by
always choosing the higher or lower returning asset each quarter.  The
historical optimal timing path is, however, indistinguishable from a
random sequence; it is unpredictable and codes no information about
the future optimal timing path.

The key observation is that return is a multiplicative process and so
its PDF is log-normal.  The implication is the mathematical fact that
the most probable outcome from market timing is a below median
return---even before accounting for costs.  This stems from an
elementary property of the logarithm.  Put another way, simple math
says the most likely outcome of market timing is under performance.
Exactly what this under performance is can be ascertained because the
median of the market timing return PDF can be directly calculated as a
weighted average of the returns of the model assets with weights given
by the fraction of time periods each asset has a higher return than
the other.  For the time period of the data the median return was
close to the return of the static 60:40 stock:bond balanced index;
althrough, the value of $p_b$ need not be fixed for all time.

For simplicity of analysis and clarity of results the model in this
paper has only two asset classes; however, it is clear that the
methodology could be extended to any number of asset classes.

\appendix

\section{Index Funds}
\label{sec:index_funds}

Fund data scrapped from Yahoo Finance on 2 November 2017.  Data covers
1993Q1 through 2017Q3 for
\begin{description}
\item{Vanguard Total Stock Market Index (VTSMX)}
\item{Vanguard Total Bond Market Index (VBMFX)}
\item{Vanguard Balanced Index (VBINX)}
\end{description}

\section{NIST Test Suite}
\label{sec:nist_details}

The NIST test suite \cite{Bassham_NIST_2010} consists of 15
statistical tests that test the randomness of binary sequences.  The
suite is designed to test the quality of random or pseudo-random
number generators, which is why for a reliable result most of the
tests require much longer (by several orders of magnitude) data
sequences than that available from the financial time series of stock
and bond returns.  However, for four of the tests a 99 bit sequence is
close to the minimum suggested length of 100 bits.  It should be noted
that two of the 15 tests test for uniformity of the binary
sequence---the equal probability of the number of ones and zeros---and
so are not applicable to the historically optimal timing sequence.  I
use Gerhardt's implementation of the test suite for Mathematica
\cite{Gerhardt_nist_2010}.

The following, taken from \cite{Bassham_NIST_2010}, gives a brief
description of the four NIST tests I use. After each description I
give the $P$-value for the test, where $P \ge 0.01$ indicates $f_b$ is
random to a 99\% confidence level.  The four tests are:
\begin{description}
\item{Runs Test.}  The purpose of the runs test is to determine
  whether the number of runs of ones and zeros of various lengths is
  as expected for a random sequence.  In particular, this test
  determines whether the oscillation between such zeros and ones is
  too fast or too slow.  $P= 0.80$.

\item{Discrete Fourier Transform (Spectral) Test.}  The purpose of
  this test is to detect periodic features (i.e., repetitive patterns
  that are near each other) in the tested sequence that would indicate
  a deviation from the assumption of randomness.  $P = 0.32$.

\item{Serial Test.}  The purpose of this test is to determine whether
  the number of occurrences of the $2^m$ $m$-bit overlapping patterns
  is approximately the same as would be expected for a random
  sequence.  $P = 0.50$.

\item{Cumulative Sums Test.}  The purpose of the test is to determine
  whether the cumulative sum of the partial sequences occurring in the
  tested sequence is too large or too small relative to the expected
  behavior of that cumulative sum for random sequences.  $P = 0.01$.
\end{description}

\section{Calculation of Expectation Value}
\label{sec:calculation_detailed}

Start with the geometric mean of equation~\ref{eq:return_path}, take
the log of both sides then apply the expection operator to obtain
\begin{equation}
\label{eq:expectation_return_path_1}
{\boldmath E}\left[\log\left[\rho^{1/N}\right]\right] =
{\boldmath E}\left[
\log\left[\left(\prod_i^N \left( f_{i} r_{si} + (1 - f_{i}) r_{bi}
    \right) \right)^{1/N} \right] \right].
\end{equation}
Denoting $\mu = {\boldmath
  E}\left[\log\left[\rho^{1/N}\right]\right]$, by linearity of the
expectation operator
\begin{equation}
\label{eq:expectation_return_path_2}
\mu = N^{-1}
  \sum_i^N {\boldmath E}\left[\log\left( f_{i} r_{si} + (1 - f_{i}) r_{bi} \right)
  \right]
\end{equation}
and by Jensen's inequality
\begin{equation}
\label{eq:expectation_return_path_3}
\mu \le N^{-1}
  \sum_i^N \log\left({\boldmath E}\left[ f_{i} r_{si} + (1 - f_{i})
      r_{bi} \right] \right).
\end{equation}
For simplicity I assume the equality holds in
equation~\ref{eq:expectation_return_path_3} in light of the supporting
numerical results.  Then by repeated use of the linearity of
${\boldmath E}$, the expression inside the log becomes
\begin{equation}
  \label{eq:expectation_return_path_4}
  \begin{aligned}
{\boldmath E}\left[ f_{i} r_{si} + (1 - f_{i}) r_{bi} \right] 
&= 
{\boldmath E}\left[ f_{i}\right]  {\boldmath E}\left[ r_{si} \right] +
{\mathrm Cov}\left[f_i, r_{si}\right] \\
&\quad +
\left(1 - {\boldmath E}\left[f_{i} \right]\right) {\boldmath E}\left[r_{bi} \right] 
+ {\mathrm Cov}\left[1 - f_i, r_{bi}\right],
\end{aligned}
\end{equation}
where ${\mathrm Cov}$ is covariance.  Since the $f_i$ are created in
this paper as random sequences, they are uncorrelated with the stock
and bond returns $r_{si}$ and $r_{bi}$, and both covariance terms are
zero; althrough, timing paths could be constructed that are
corrrelated with returns, in which case the covariance terms would be
non-zero.  By construction ${\boldmath E}\left[ f_{i}\right] = p_b$.
And, after noting that
${\boldmath E}\left[ r_{si} \right] = \bar{r}_{s} $ and
${\boldmath E}\left[ r_{bi} \right] = \bar{r}_{b}$ are the geometric
means of respectively the stock and bond returns,
equation~\ref{eq:evalue_log_geo_mean} results.

\section{Tactical Allocation Funds}
\label{sec:taa_funds}

Fund data and objectives summaries scrapped from Yahoo Finance 6
November 2017.  Data covers 1994Q1 through 2017Q3.  I do not mean to
pick on these funds.  They were the only ones in Morningstar
that were both in the tactical asset allocation
category and had a sufficiently long, publicly available return
series.  As Morningstar rates both funds four out of five stars, and
they have been in business 23 years, and they have billions of dollars
of assets under management, clearly these funds are above average,
successful market timing funds.  Only, they did not produce above
average returns for their investors.
\begin{description}
\item{Putnam Dynamic Asset Allocation Balanced (PABAX).}  The fund
  allocates 45\% to 75\% of its assets in equities and 25\% to 55\% in
  fixed income securities. It invests mainly in equity securities
  (growth or value stocks or both) of both U.S. and foreign companies
  of any size. The fund also invests in fixed-income investments,
  including U.S. and foreign government obligations, corporate
  obligations and securitized debt instruments (such as
  mortgage-backed investments).
  
\item{Putnam Dynamic Asset Allocation Conservative (PACAX).}  The fund
  allocates 15\% to 45\% of its assets in equities and 55\% to 85\% in
  fixed income securities. It invests mainly in fixed-income
  investments, including U.S. and foreign government obligations,
  corporate obligations and securitized debt instruments. The fund
  also invests, to a lesser extent, in equity securities (growth or
  value stocks or both) of U.S. and foreign companies of any size.
\end{description}

\bibliographystyle{unsrt}
\bibliography{timing}

\end{document}